\documentclass{aa}
\usepackage{epsf}

\begin{document}


\newcommand{\pF}{\mbox{$p_{\mbox{\raisebox{-0.3ex}{\scriptsize F}}}$}}  
\newcommand{\M}{\mbox{$\dot{M}_{10}$}}
\newcommand{\vph}[1]{\mbox{$\vphantom{#1}$}}  
\newcommand{\kB}{\mbox{$k_{\rm B}$}}           
\newcommand{\vF}{\mbox{$v_{\mbox{\raisebox{-0.3ex}{\scriptsize F}}}$}}  
\renewcommand{\arraystretch}{1.5}
\newcommand{\dd}{{\rm d}}


\title{Thermal State of Transiently Accreting Neutron Stars}
\author{
       D.~G.~Yakovlev\inst{1}
\and
       K.~P.~Levenfish\inst{1}
\and
       P.~Haensel\inst{2}
}
\institute{
        Ioffe Physical Technical Institute, Politekhnicheskaya 26,
        194021 St.-Petersburg, Russia \\
\and
       Copernicus Astronomical Center,
       Bartycka 18, 00-716 Warsaw, Poland\\
        {\it e-mails: yak@astro.ioffe.ru,
                      ksen@astro.ioffe.ru,
                      haensel@camk.edu.pl}
}

\date{}
\offprints{D.\ Yakovlev: yak@astro.ioffe.ru}

\titlerunning{Thermal state of transiently accreting neutron stars}
\authorrunning{D.~Yakovlev et al.}

\abstract{We study
thermal states of transiently accreting neutron
stars (with mean accretion rates $\dot{M} \sim 10^{-14}-10^{-9}$
M$_\odot$ yr$^{-1}$) determined by the deep crustal heating of accreted matter
sinking into stellar interiors.
We formalize a direct correspondence of this problem
to the problem of cooling neutron stars. 
Using a simple toy model
we analyze the most important factors which affect the thermal states of
accreting stars:
a strong superfluidity in the cores of low-mass stars
and a fast neutrino emission
(in nucleon, pion-condensed, kaon-condensed, 
or quark phases of dense matter)
in the cores of high-mass stars. We briefly compare the results
with the observations of soft X-ray transients in quiescence.
If the upper limit on the quiescent thermal luminosity of 
the neutron star in SAX J1808.4--3658 (Campana et al.\
\cite{campanaetal02}) is associated with the deep crustal heating, 
it favors the model of nucleon neutron-star cores
with switched-on direct Urca process.
}

\maketitle


\section{Introduction}

Great progress in observations of soft X-ray transients (SXRTs)
in quiescence has attracted attention
to these objects. We consider
the SXRTs containing neutron stars (NSs)
in binary systems with low-mass companions
(low-mass X-ray binaries); see Chen et al.\ (\cite{csl97})
for a review.
They undergo the periods of outburst activity 
(days--months, sometimes years)
superimposed with
the periods of quiescence (months--decades). This transient activity is
regulated most probably
by the regime of accretion from the disks around the NSs. 
An active period is associated with a switched-on
accretion; the accretion energy released at the
NS surface is high enough for a transient
to look like a bright X-ray source
($L_X \sim 10^{36}-10^{38}$
erg s$^{-1}$). The accretion is
switched off or strongly suppressed during quiescence intervals when
the NS luminosity drops by several orders of
magnitude ($L_X \la 10^{34}$
erg s$^{-1}$). 

The nature of the quiescent emission is still a subject
of debates. The hypothesis that this emission
is produced by the thermal flux emergent from the NS
interior has been rejected initially due to
two reasons. First, the radiation spectra
fitted by the blackbody model have given unreasonably
small NS radii. Second, NSs in SXRTs have been expected
to be internally cold; their quiescent
emission should have been much lower than the observed one. 
These arguments were questioned by Brown et al.\ (\cite{bbr98}). 
They suggested that the NSs can be warmed up to the
required level by
{\it the deep crustal heating}
associated with nuclear transformations
in accreted matter sinking in the NS interiors
(Haensel \& Zdunik \cite{hz90}),
while the radiation spectra can be fitted with the NS hydrogen
atmosphere models (with realistic NS radii).
It turned out that 
the emergent radiation flux may depend on the NS internal structure
which opens an attractive possibility to explore
the internal structure and
the equation of state of dense matter by comparing
the observations of SXRTs in quiescence
with theoretical models (e.g., Ushomirsky \& Rutledge \cite{ur01},
Colpi et al.\, \cite{cgpp01},
Rutledge et al.\ \cite{rbbpzu02},
Brown et al.\ \cite{bbc02}, and references therein).

In this article, we discuss this possibility in more detail
by making use of the close relationship
between the theory of thermal states of transiently accreting NSs
and the theory of cooling isolated NSs.
Since the equation of state of NS matter is still poorly known
we will make a general 
analysis of the problem with a simple toy model of NS  
thermal structure (described by Yakovlev \& Haensel \cite{yh02}).
It will enable us to  
confront the theory with the observations
without performing complicated calculations. 
 
\section{Numerical simulations}

Let us study a thermal state of a transiently accreting
NS. Since the quiescence intervals
are much shorter than typical time scales of
NS thermal relaxation 
($\sim10^4$ yrs, Colpi et al.\ \cite{cgpp01})
we neglect short-term variability
in the NS crust; it can be associated
with variable residual accretion in quiescence, thermal
relaxation of transient 
deep crustal heating, etc.; see, e.g.,
Ushomirsky \& Rutledge (\cite{ur01}) and
Brown et al.\ (\cite{bbc02}). 
Instead, we focus on a (quasi)stationary
steady state of the NS
determined by the mass
accretion rate $\dot{M}\equiv \langle \dot{M} \rangle$ 
(from $10^{-14}$ to $10^{-9}$ M$_\odot$ yr$^{-1}$)
averaged over time intervals comparable with the thermal
relaxation time scales.
The accretion rates of study are too low 
to noticeably increase NS mass, $M$, 
during long periods of the NS evolution.
An accreted material sinks gradually into the deep
layers of the NS crust (into the density range
from $\sim 10^{10}$ g cm$^{-3}$ to $\sim 10^{13}$ g cm$^{-3}$), 
where it undergoes various
transformations (beta captures, neutron emissions and
absorptions, pycnonuclear reactions) accompanied by
a substantial energy release. This produces
the deep crustal heating whose power
is determined by $\dot{M}$. 
Complete sequence of nuclear transformations
and associated energy release
was calculated by Haensel \& Zdunik (\cite{hz90})
(see Bisnovatyi-Kogan \cite{bisnovatyi}, for references 
to some earlier work).
The total energy release is about 1.45 MeV per 
accreting nucleon. 
The main energy release
takes place at densities from about $10^{12}$
to $10^{13}$ g cm$^{-3}$, about 1 km under the
surface.
The heating power is estimated
as 
\begin{equation}
  L_{\rm dh} = 1.45~{\rm MeV} \, \dot{M}/m_{\rm N}
  \approx 8.74 \times 10^{33}  \, \M\;\;{\rm   erg\; s}^{-1},
\label{Ldh} 
\end{equation}
where $\M \equiv \dot{M}/(10^{-10} \, {\rm M}_\odot \;{\rm yr}^{-1})$
and $m_{\rm N}$ is the nucleon mass.
Independently of the
initial thermal state (cold or hot), a NS will eventually
reach a stationary thermal state supported by this heating.

We stress that $L_{\rm dh}$ is much smaller than the
mean energy release rate at the NS surface (where
kinetic accretion energy transforms into heat) and in the outermost surface
layers (at densities $\rho \la 10^{10}$ g cm$^{-3}$, where
the accreted hydrogen or helium burns 
into heavier elements). 
We expect that the energy 
released at $\rho \la 10^{10}$ g cm$^{-3}$ is radiated away by the surface
photon emission and does not heat the NS interior.
This assumption is especially true for transiently
accreting NSs, where the surface heat 
is rapidly emitted
from the surface at the beginning of quiescent stages.

The deep-heating theory 
of Haensel \& Zdunik (\cite{hz90}) assumes that the 
burning in the very surface layers  
leads to the production of iron-like elements at $\rho \sim 10^{10}$
g cm$^{-3}$. According to the recent studies
by Schatz et al.\ (\cite{schatzetal01})
thermonuclear burning of hydrogen in
the surface layers of accreting NSs may end at much heavier
elements such as Sn, Sb and Te. 
The associated deep heating has been most recently
analyzed by Haensel \& Zdunik (\cite{hz03})
with the result that the heavy elements in the surface
layers do not significantly affect the composition of accreted
matter and the energy release in the deeper layers of the inner crust 
(where the pycnonuclear reactions proceed);
the total energy release remains about (1.1--1.5) MeV
per nucleon.    

To study the thermal states of accreting NSs we employ
a simple toy model of cooling isolated NSs
described by Yakovlev \& Haensel (\cite{yh02}).
We have tested the results with those obtained with the
``exact'' codes which calculate the thermal states of
cooling NSs and accreting NSs; the agreement has turned
out to be satisfactory.
Taking into account many theoretical
and observational uncertainties,
the toy model seems sufficient at the present
stage of the investigation.

The toy model assumes that the NS core
is divided into three zones:
the {\it outer} zone, $\rho< \rho_{\rm s}$;
the {\it transition} zone, $\rho_{\rm s} \leq \rho < \rho_{\rm f}$;
and the {\it inner} zone, $\rho \geq \rho_{\rm f}$.
If the NS central density
$\rho_{\rm c} \leq \rho_{\rm s}$, two last zones are absent.

In the outer zone, the neutrino emission is supposed to be
{\it slow}, while in the inner zone it is
{\it fast}. The neutrino emissivity
$Q_\nu$ (erg s$^{-1}$ cm$^{-3}$) is assumed to be:
\begin{equation}
     Q_\nu^{\rm slow}(\rho \leq \rho_{\rm s})=Q_{\rm s} T_9^8,\qquad
     Q_\nu^{\rm fast}(\rho \geq \rho_{\rm f})=Q_{\rm f} T_9^6.
\label{Qs}
\end{equation}
Here, $T_9$ is the internal stellar temperature $T$
expressed in $10^9$ K, while $Q_{\rm s}$ and $Q_{\rm f}$
are constants.
For simplicity, the toy model uses the linear interpolation in $\rho$ between
$Q_\nu^{\rm slow}$ and $Q_\nu^{\rm fast}$ in the transition zone.
The approximation is crude but sufficient
for exploring the main unknown features of
the transition zone: its position and thickness.

This {\it generic} description of $Q_\nu$
covers many {\it physical} models
of nucleon and exotic supranuclear
matter with different leading neutrino processes
listed in Tables 1 and 2 
(from Yakovlev \& Haensel \cite{yh02} with the kind
permission of the authors). In these tables,
N is a nucleon (neutron or proton, n or p); e is an electron;
$\nu$ and $\bar{\nu}$ are neutrino and antineutrino;
q is a quasinucleon (mixed n and p states); while u and d are quarks.

In particular, $Q_{\rm s}$ describes the
modified Urca (Murca) process
in a nonsuperfluid nucleon matter,
or weaker
NN-bremsstrahlung (e.g., nn-bremsstrahlung if Murca is
suppressed by a strong proton superfluidity
as considered by Kaminker et al.\ \cite{kyg02}).
The factor $Q_{\rm f}$ describes the processes
of fast neutrino emission:
a very powerful direct Urca (Durca) process in nucleon matter
(Lattimer et al.\ \cite{lpph91}) or
somewhat weaker similar
processes in exotic phases of
matter (pion condensed, kaon condensed,
or quark matter) as reviewed, e.g., by Pethick (\cite{pethick92}).
If hyperons appear in the NS core in addition to nucleons,
the neutrino emissivity is expected to be qualitatively
the same as in the nucleon core.
The bottom line of Table 2
corresponds to a nonsuperfluid quark matter in NS cores.

\newcommand{\rrr}{\rule{0cm}{0.2cm}}

\begin{table}[t]
\caption{Main processes of slow neutrino emission
in nucleon matter: Murca and
bremsstrahlung (brems)}
\begin{center}
  \begin{tabular}{|lll|}
  \hline
  Process   &    &  $Q_{\rm s}$, erg cm$^{-3 \rrr}$ s$^{-1}$ \\
  \hline
  Murca &
  ${\rm nN \to pN e \bar{\nu} \quad
   pN e \to nN \nu } $ &
  $\quad 10^{20 \rrr}-3 \times 10^{21}$  \\
  Brems. &
  ${\rm NN \to NN  \nu \bar{\nu}}$  &
  $\quad 10^{19 \rrr}-10^{20}$\\
   \hline
\end{tabular}
\label{tab-nucore-slow}
\end{center}
\end{table}

\begin{table}[t]
\caption{Leading processes of fast
neutrino emission
in nucleon matter and three models of exotic
         matter}
\begin{center}
  \begin{tabular}{|lll|}
  \hline
  Model              & Process             &
        $Q_{\rm f}$, erg cm$^{-3 \rrr}$ s$^{-1}$ \\
  \hline
  Nucleon matter &
  ${\rm n \to p e \bar{\nu} \quad
   p e \to n \nu }$ & $\quad
  10^{26 \rrr}-10^{27}$  \\
  Pion condensate &
  ${\rm q \to q e \bar{\nu} \quad
   q e \to q {\nu} } $ & $ \quad
  10^{23 \rrr}-10^{26}$  \\
   Kaon condensate &
   ${\rm q \to q e \bar{\nu} \quad
    q e \to q \nu } $ & $ \quad
   10^{23 \rrr}-10^{24}$ \\
   Quark matter &
   ${\rm d \to u e \bar{\nu} \quad  u e \to d \nu } $ & $  \quad
   10^{23 \rrr}-10^{24}$ \\
   \hline
\end{tabular}
\label{tab-nucore-fast}
\end{center}
\end{table}

The transition zone models a switch-on of the fast neutrino
emission with growing $\rho$. 
In a nonsuperfluid matter, the lower density
$\rho_{\rm s}$  
is a threshold density of the fast
emission; this threshold is usually sharp, i.e.,
$\rho_{\rm s} \approx \rho_{\rm f}$.
The superfluidity suppresses the neutrino emission.
Even if the superfluidity strength is high at the formal
(nonsuperfluid) threshold, it decreases with growing $\rho$
in all realistic microphysical models of superfluid matter.
In this case, $\rho_{\rm s}$ is the density where
the superfluid suppression of the fast neutrino emission
becomes weaker, while $\rho_{\rm f}$ is the density
where this suppression almost vanishes.
Thus the superfluidity can broaden a switch-on
of the fast neutrino emission and produce
a sufficiently wide transition zone.
These effects are explained, e.g.,
by Yakovlev et al.\ (\cite{ygkp02}), taken nucleon
Durca process and strong proton superfluidity as an example.
In addition to reducing the ordinary neutrino 
emission processes, 
the superfluidity initiates a specific neutrino emission
associated with Cooper pairing of baryons
(Flowers et al.\ \cite{frs76}). We neglect this emission
in the present calculations but outline its consequences in Sect.\ 4.

The toy model solves the equation of thermal balance
of a cooling NS in the approximation of isothermal interior
(see below) under a number of simplified assumptions
on NS internal structure (Yakovlev \& Haensel \cite{yh02}).
The density profile in the star is approximate,
$\rho(r)=\rho_{\rm c}\, ( 1 - r^2/R^2)$, so that the NS mass
is $M=8 \pi R^3 \rho_{\rm c}/15$, 
$R$ being the stellar radius.
The stiffness of various
equations of state can be mimicked by choosing different
$M-R$ relations. For simplicity, following
Yakovlev \& Haensel (\cite{yh02}), 
we set $R=12$ km and vary $\rho_{\rm c}$
from $7 \times 10^{14}$ to $1.4 \times 10^{15}$ g cm$^{-3}$,
varying thus $M$ from 1.02 M$_\odot$ to 2.04 M$_\odot$.
More realistic $M-R$ relations will not change our
principal conclusions. Various cooling regimes are
regulated by four parameters: $Q_{\rm f}$,
$Q_{\rm s}$, $\rho_{\rm s}$, and $\rho_{\rm f}$,
and are studied (Sect.\ 4) without complicated computation.

We have updated the toy model by incorporating
the deep crustal heating as described below.

\section{Relation to cooling of isolated stars}
 
A thermal state of an accreting NS
is similar to a state of a cooling isolated NS
with isothermal interior. 

\begin{figure}
\centering
\epsfxsize=8.7 cm
\epsffile[30 160 560 620]{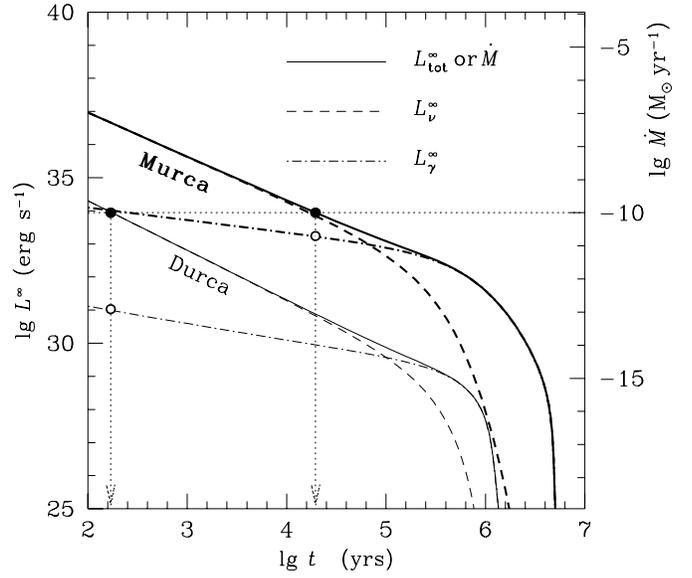}
\caption{
Simultaneous calculation of cooling and heating curves
for low-mass (thick lines) and high-mass 
(thin lines) NS models (see the text). Displayed are: the neutrino, photon
and total luminosities, $L_\nu^\infty$, $L_\gamma^\infty$ 
and $L_{\rm tot}^\infty$
(left vertical scale), of a cooling NS
versus its age, and the accretion rate $\dot{M}$ (right
vertical scale) of an accreting NS 
star with $L_{\rm dh}^\infty=L_{\rm tot}^\infty$.  
Dotted horizontal line: $L_{\rm dh}^\infty$ for $\dot{M}=10^{-10}$
M$_\odot$ yr$^{-1}$. Filled circles and dotted vertical arrows
refer to those low-mass and high-mass
cooling NSs which are equivalent to accreting NSs with 
the indicated $\dot{M}$; open circles: $L_\gamma^\infty$
for these cooling NSs.
}
\label{fig1}
\end{figure}

Both states can be found by solving the
equation of thermal balance of
a NS with isothermal interior 
(e.g., Glen \& Sutherland \cite{gs80}):
\begin{equation}
   C(T_i) \, {{\rm d}T_i \over {\rm d}t}=L_{\rm dh}^\infty(\dot{M})
       -L_\nu^\infty(T_i)-L_\gamma^\infty(T_{\rm s}),
\label{therm-balance}
\end{equation}
where $T_{\rm s}$ is the effective surface temperature,
$T_i(t)=T(r,t) \, {\rm e}^\Phi$ is the redshifted internal
temperature which is constant throughout the isothermal
interior ($\rho \ga \rho_{\rm b} \sim 10^{10}$ g cm$^{-3}$) 
with account for the
effects of General Relativity; $T(r,t)$ is the local
internal temperature of matter, and $\Phi(r)$ is the metric
function. The relation between $T_i$ and $T_{\rm s}$ 
is known from the solution of the thermal conduction
problem in the outermost heat-blanketing 
stellar envelope ($\rho < \rho_{\rm b}$).
Furthermore, $C$ is the total heat capacity
of the star, $L_\gamma^\infty=4 \pi \sigma T_{\rm s}^4 \, R^2 \,
(1-r_{\rm g}/R)$ is the photon surface luminosity
as detected by a distant observer ($r_{\rm g}=2GM/c^2$
being the gravitational radius), 
and
\begin{equation}
    L^\infty_{\rm \nu, dh}= 4 \pi \int_0^R {{\rm d}r \; r^2 \, 
      Q_{\rm \nu, dh} \,{\rm e}^{2 \Phi} \over
      \sqrt{1-2Gm/(c^2 r)}},
\label{L}
\end{equation}
are the redshifted neutrino luminosity and
the redshifted power of the deep crustal heating;
$Q_\nu$ is the neutrino emissivity; $Q_{\rm dh}$ 
is the energy release rate due to the deep crustal heating,
and $m(r)$ is the gravitational mass contained in
a sphere with radial coordinate $r$. In addition, we
introduce 
the effective surface temperature detected by a distant
observer, $T_{\rm s}^\infty=T_{\rm s} \; \sqrt{1-r_{\rm g}/R}$, 
and the ``apparent'' NS radius, $R^\infty=R / \sqrt{1-r_{\rm g}/R}$. 
Then
$L_\gamma^\infty=4 \pi \sigma (T_{\rm s}^\infty)^4 \, (R^\infty)^2$.

\begin{figure}
\centering
\epsfysize=7.5 cm
\epsffile[80 215 544 670]{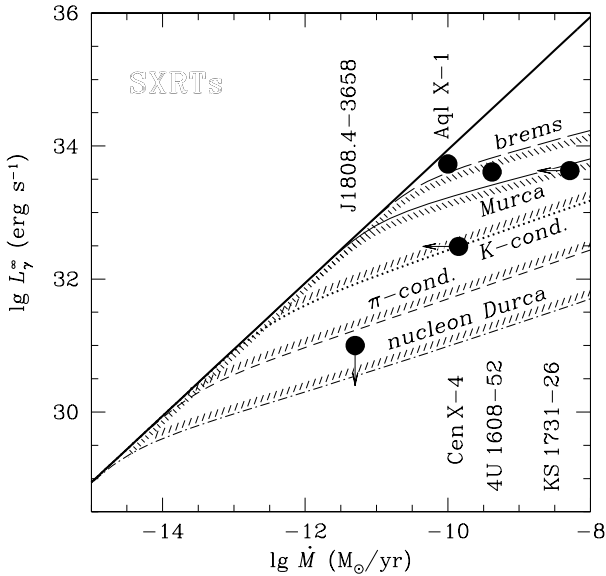}
\caption{
Heating curves of accreting NSs
compared with observations of several SXRTs in quiescence.
Thick solid line: deep crustal heating power, $L_{\rm dh}$.
Long dashes and thin solid lines: two models of
heating curves of low-mass NSs ($Q_{\rm s}=3 \times 10^{19}$
and $10^{21}$). Dotted, short-dashed, and dot-and-dashed
lines: three models of high-mass NSs 
($Q_{\rm f}=10^{23}$, $10^{25}$, and $10^{27}$). 
}
\label{fig3}
\end{figure}

As mentioned above, we replace {\it a transient accretion} by
{\it a steady-state accretion} with the mean
constant accretion rate $\dot{M}=\langle \dot{M} \rangle$. 
We will study thus the steady states
of NSs accreting at constant rates.
A steady-state accretion in General Relativity
is characterized by a constant mass accretion rate $\dot{M}$
which determines constant number of accreting baryons
passing through a sphere of any radial coordinate $r$
per unit time for a distant observer
(e.g., Thorne \& Zytkow \cite{tz77}). 
Gravitational redshift in a thin
NS crust may be regarded as constant. In this approximation,
sufficient for our problem, we have constant
${\rm e}^\Phi = \sqrt{1-r_{\rm g}/R}$
throughout the crust, and $L_{\rm dh}^\infty = L_{\rm dh} \, 
\sqrt{1-r_{\rm g}/R}$, where $L_{\rm dh}$ is given by
Eq.\ (\ref{Ldh})
(neglecting the effects
of General Relativity). In principle, 
the gravitational redshift
has to be included into $L_{\rm dh}^\infty$.
However, since $\dot{M}$ in SXRTs is determined
with large uncertainties, it is premature 
to include such effects, and we 
set $L_{\rm dh}^\infty = L_{\rm dh}$ in subsequent
calculations.

Following the cooling of an isolated NS we calculate
a {\it cooling curve}, the dependence of
the surface temperature  
or the photon surface luminosity 
on NS age, $T_{\rm s}^\infty(t)$ or $L_\gamma^\infty(t)$. 
In about
$t \sim 100$ yrs (when $T_{\rm s}^\infty$ 
drops below $3 \times 10^6$ K)
the NS interior becomes isothermal. Subsequent cooling
can be considered as a sequence
of quasistationary states. It is described
by Eq.\ (\ref{therm-balance}) with $L_{\rm dh}^\infty=0$.
The examples are presented in Fig.\ \ref{fig1} described below.

Following the evolution of an accreting NS from 
Eq.\ (\ref{therm-balance}), we would find that the NS,
starting from an arbitrary initial thermal state,
eventually reaches a stationary state supported
by the deep crustal heating (whose power, $L^\infty_{\rm dh}$, 
is given by Eq.\ (\ref{Ldh})). The steady state is reached when
the total NS luminosity, $L^\infty_{\rm tot}=L_\nu^\infty
+ L_\gamma^\infty$, is balanced by $L^\infty_{\rm dh}$. 
This steady state is described by Eq.\ (\ref{therm-balance})
in the stationary limit:
\begin{equation}
     L_{\rm dh}^\infty(\dot{M})=L_\nu^\infty(T_i) 
    + L_\gamma^\infty(T_{\rm s}),
\label{Main}
\end{equation}
where $L_{\rm dh}^\infty$ is known once $\dot{M}$
is specified. Since we are interested in the steady state,
it is sufficient to solve Eq.\ (\ref{Main}). The solutions
give us a {\it heating curve}, the dependence of the surface
temperature or the photon luminosity  
on the mean accretion rate,
$T^\infty_{\rm s}(\dot{M})$ or $L^\infty_\gamma(\dot{M})$.
For the values of $\dot{M} \la 10^{-9}$ M$_\odot$ yr$^{-1}$, 
we obtain the thermal states with 
$T_{\rm s} \la 3 \times 10^6$ K.
The examples are presented in Fig.\ \ref{fig3} discussed below.

Comparing the equations of NS cooling and accretion heating
we see that any steady state of a NS 
accreting at a rate $\dot{M}$ is equivalent
to a state of a cooling NS
in such a moment of time $t$ when
$L_{\rm tot}^\infty(t)=L_{\rm dh}^\infty(\dot{M})$.
In other words, calculating a cooling curve,
we find $L_{\rm tot}^{\infty}(t)$,
$L_\gamma^\infty(t)$ and equivalent $\dot{M}(t)$ for any moment
of time. Considering $t$ as a parameter, we immediately
get a heating curve $L_\gamma^\infty(\dot{M})$.

This procedure is illustrated in Fig.\ \ref{fig1}.
It shows a simultaneous solution of the cooling and heating
problems with the toy model (Sect.\ 2) 
which calculates the cooling of isolated low-mass
and high-mass NSs. The masses are
1.16 and 2.04 M$_\odot$ ($\rho_{\rm c}=8 \times 10^{14}$
and $1.4 \times 10^{15}$ g cm$^{-3}$). The 
toy-model parameters are:
$Q_{\rm s}=10^{21}$ erg cm$^{-3}$ s$^{-1}$
and $Q_{\rm f}=10^{27}$ erg cm$^{-3}$ s$^{-1}$
(appropriate to the Murca and Durca processes in nonsuperfluid matter), 
$\rho_{\rm s}=8 \times 10^{14}$ g cm$^{-3}$, and
$\rho_{\rm f}=10^{15}$ g cm$^{-3}$. 
The figure shows the decrease of $L_\gamma^\infty$, 
$L_\nu^\infty$, and
$L_{\rm tot}^\infty$ (left vertical axis) with time for cooling stars.
The right vertical axis presents $\dot{M}$ for accreting
NSs with $L_{\rm tot}=L_{\rm dh}$.
For instance, taking $\dot{M}=10^{-10}$ M$_\odot$ yr$^{-1}$ (horizontal
dotted line) and locating the points, where $L_{\rm tot}=L_{\rm dh}$
(filled circles and vertical dotted arrows), we find
that the low-mass NS accreting with the given $\dot{M}$ is equivalent
to the cooling NS of age $t \approx 2 \times 10^4$ yrs,
while the high-mass accreting NS is equivalent to
the cooling NS of age $t \approx 170$ yrs.
The thermal states of low-mass and high-mass accreting NSs
are seen to be drastically different
(the appropriate photon luminosities
$L_\gamma^\infty$ are marked by open circles).

It is clear that any steady state 
of an accreting NS is independent of the heat capacity
of matter and the thermal conductivity in the isothermal
NS interior, although the relaxation to this
state does depend on these quantities.
Accordingly, superfluidity in NS interiors affects the steady
state only by regulating the neutrino emission processes.

Although a general analysis of Eq.\ (\ref{Main})
is complicated, we can mention two limiting regimes.

(i) {\it The photon emission regime} is realized 
in cold enough NSs, where $L_\nu^\infty \ll L_\gamma^\infty$,
i.e., the heat released in the deep crust is carried
away by the thermal conductivity to the surface and emitted 
in the form of the surface photon radiation.
This regime is equivalent to the photon emission stage
of cooling NSs. In this regime, Eq.\ (\ref{Main})
reduces to: $L^\infty_{\rm dh}(\dot{M})=L_\gamma^\infty(T_{\rm s})$,
which immediately gives an estimate 
$T_{\rm s6} \equiv T_{\rm s}/10^6~{\rm K} \sim 2 \dot{M}_{10}^{1/4}$ 
(for $R \sim 10$ km).
Accordingly, the surface temperature is determined
by the accretion rate, and it is independent of the internal
structure of the NS. The internal temperature
can be found then using the $T_{\rm s}-T_i$ relation.
For typical surface gravities $g_{\rm s} \sim 2 \times 10^{14}$
cm s$^{-2}$ from the relation of Gudmundsson et al.\
(\cite{gpe83}) we have $T_{i9}
\equiv T_i/10^9~{\rm K} \sim 0.1 T_{\rm s6}^2$. 
Thus, $T_{i9} \sim 0.4 \sqrt{\M}$.

(ii) {\it The neutrino emission regime} is realized
in warmer NSs, where $L_\nu^\infty \gg L_\gamma^\infty$,
i.e., the heat released in the deep crust is spread
by the thermal diffusion over the star and carried away
by the neutrino emission. This regime is equivalent to
the neutrino stage in cooling NSs. The
thermal balance equation then reads: $L^\infty_{\rm dh}(\dot{M})=
L_\nu^\infty(T_i)$, which gives $T_i$.
The surface temperature can be found then using
the $T_{\rm s}-T_i$ relation. In this regime,
the surface temperature $T_{\rm s}$ does depend on
the internal structure of the star, on the equation of state,
the presence or absence of enhanced neutrino emission
mechanism, and on the nature of this emission.
For the assumptions implied in our toy model, 
we have two distinctly different
cases.

(ii.a) {\it In low-mass NSs} ($\rho_{\rm c} \leq \rho_{\rm s}$)
we have slow neutrino emission, 
$Q_\nu^{\rm slow}\!=Q_{\rm s}T^8_{i9}$.
Introducing $Q_{\rm s21} \equiv Q_{\rm s}/(10^{21}$ erg s$^{-1}$ cm$^{-3}$),
we have $L_\nu \sim 5 \times 10^{39}T_{i9}^8\,Q_{\rm s21}$
erg s$^{-1}$.
Then we obtain $T_{i9} \sim 0.18\,  
(\M /Q_{\rm s21})^{1/8}$,
and $T_{\rm s6} \sim 1.4 \, (\M/Q_{\rm s21})^{1/16}$;
the surface temperature weakly depends
on the accretion rate and the neutrino emission level $Q_{\rm s}$
(as a result of the strong temperature
dependence of $L_\nu$). The comparison with case (i)
shows that the transition from the photon emission regime
to the neutrino emission regime takes place
at $\M \sim 0.13/Q_{\rm s21}^{1/3}$ and $T_{\rm s6} \sim 
1.2/Q_{\rm s21}^{1/12}$.  

(ii.b) {\it In high-mass NSs} ($\rho_{\rm c}\!>\! \rho_{\rm f}$)
the neutrino~emission is mainly produced by the fast
mechanism,
$Q_\nu^{\rm fast}\!=Q_{\rm f}T^6_{i9}$, 
in the inner
zone ($\rho\!>\!\rho_{\rm f}$) of the NS core. 
Introducing $Q_{\rm f27} \equiv Q_{\rm f}/(10^{27}$ erg s$^{-1}$
cm$^{-3}$), 
we have   
$L_\nu \sim 5 \times 10^{45}T_{i9}^6 \,Q_{\rm f27}$
erg s$^{-1}$,
 $T_{i9} \sim 0.01\, (\M /Q_{\rm f27})^{1/6}$,
and $T_{\rm s6} \sim 0.3 \, (\M/Q_{\rm f27})^{1/12}$;
$T_{\rm s}$ is again a weak function of $\dot{M}$ and $Q_{\rm f}$.
A comparison with case (i)
shows that the transition between the photon and neutrino emission 
regimes occurs
at $\M \sim 10^{-5}/\sqrt{Q_{\rm f27}}$ and 
$T_{\rm s6} \sim 0.1/Q_{\rm f27}^{1/8}$.  
We have a much colder NS than in case (ii.a),
and the neutrino emission regime persists to much lower $\dot{M}$.

\section{Results and discussion}

Our numerical calculations with the toy model
confirm this qualitative analysis. 
The results are presented in Fig.\ \ref{fig3} which
shows the surface photon luminosity versus
accretion rate for low-mass and high-mass NSs.

The thick solid curve presents the deep heating power, $L_{\rm dh}^\infty$,
which is the upper limit of $L_\gamma^\infty$ for any accreting source. 

Moving from top to bottom, the next two lines refer
to low-mass NSs with two types of slow neutrino emission
appropriate either to neutron-neutron bremsstrahlung
in the NS cores with a strong proton superfluidity 
($Q_{\rm s}= 3 \times 10^{19}$) or
to the Murca process in nonsuperfluid cores ($Q_{\rm s}=10^{21}$).
The strong proton superfluidity 
damps the Murca process
and enables us to obtain hotter
NSs, just as in the theory of cooling NSs (e.g., Kaminker et al.\
\cite{kyg02}).

The three next lines refer
to high-mass NSs with three types
of fast neutrino emission appropriate to (from top to bottom)
kaon-condensed matter ($Q_{\rm f}=10^{23}$),
pion-condensed matter ($Q_{\rm f}=10^{25}$),
or nucleon matter with open Durca process ($Q_{\rm f}=10^{27}$).
NSs with hyperons cores are expected to cool at about
the same rate as NSs with nucleon cores.

The heating curves of low-mass NSs provide the upper 
limit of $L_\gamma^\infty$, while the curves of high-mass
stars give the lower limit of $L_\gamma^\infty$, for
any particular equation of state (EOS) of NS interiors
(for any set of the four neutrino emission parameters,
in our case). Accordingly, the upper limits are shaded below
the curves, and the lower limits are shaded above the curves.
As in the cooling theory (Yakovlev \& Haensel
\cite{yh02}), the upper and lower heating curves
are almost insensitive to the position and width of
the transition zone if $\rho_{\rm f}$ and $\rho_{\rm s}$
are located anywhere between $8 \times 10^{14}$ and
$1.2 \times 10^{15}$ g cm$^{-3}$.
The upper curve is determined by the parameter $Q_{\rm s}$,
while the lower one is determined by $Q_{\rm f}$.

Varying the NS mass from the lowest values to the highest
we obtain a family of heating curves which fill in the space
in Fig.\ \ref{fig3} between the upper and lower curve
for a given EOS of dense matter.
The group of NSs whose heating curves lie essentially
between the upper and lower curves will be called {\it medium-mass}
stars. Their central density falls within the transition layer
between the slow and fast neutrino
emission zones ($\rho_{\rm s} \la \rho
\la \rho_{\rm f}$, Sect.\ 2), and their mass range 
is sensitive to the position and
width of this layer.  The same situation occurs in the theory
of cooling NSs (e.g., Yakovlev \& Haensel \cite{yh02}).
Thus, for a given EOS of dense matter we
obtain its own upper and lower heating curves,
and intermediate heating curves of medium-mass stars.

In analogy with the
cooling theory, the existence of a representative class
of medium-mass stars (a smooth transition from
the upper to the lower heating curves with increasing $M$)
depends on the relative width of the transition
zone, $\Delta \rho / \rho_{\rm s} \equiv 
(\rho_{\rm f}-\rho_{\rm s})/\rho_{\rm s}$, and on the contrast
between the fast and slow neutrino emissivities, $Q_{\rm f}/Q_{\rm s}$.
To ensure the existence of this NS class
for a sharp emissivity contrast, $Q_{\rm f}/Q_{\rm s} \gg 10^3$,
we need
a rather wide transition zone, $\Delta \rho/\rho_{\rm s} \ga 0.1$.
On the other hand,
this class will be available even for a negligibly narrow
zone if the emissivity contrast is lower, $Q_{\rm f}/Q_{\rm s} \la 10^3$.
  
These results can be confronted with the observations of SXRTs
containing NSs. Although our theoretical toy model
is oversimplified, in Fig.\ \ref{fig3} 
we present an example of such an analysis
for five SXRTs: Aql X-1, Cen X-4, 4U 1608--522, 
KS 1731--26, and SAX J1808.4--3658.
The data are rather uncertain. Thus we plot the
observational points as thick dots.

\begin{table*}[ht]
\caption{Parameters of NSs in SXRTs 
}
\begin{center}
\begin{tabular}{|l|l||r|c|c|ll|}
\cline{1-2}
\cline{3-7}
Source   
         & $\dot{M}$, ${\rm M}_\odot$ yr$^{-1}$
         & $L_\gamma^\infty$, erg s$^{-1}$ 
         & $T_s^\infty$, eV
	 & $R^\infty$, km
	 & \multicolumn{2}{|c|}{ Reference} \\
\cline{1-2}
\cline{3-7}
Aql X-1  
         & $1.0\times10^{-10}$
         & $5.3\times 10^{33}$
	 & 113 
	 & 15.9
	 & Rutledge et al.\ (\cite{rbbpz02}) :
	 & Table 6, fit 10 \\
Cen X-4  
         & $1.4\times 10^{-10}$ 
         & $3.1\times 10^{32}$ 
	 & $\;$76$^{\ast )}$  
	 & 12.9  
	 & Rutledge et al.\ (\cite{rbbpz01}) :
	 & Table 4 \\
4U 1608--522 
         & $4.2\times 10^{-10} $ 
	 & $4.1\times 10^{33} $ 
	 & $\;\:$170$^{\ast )}$ 
	 & 9.4 
	 & Rutledge et al.\ (\cite{rbbpz99}) :
	 & Table 2 \\
KS 1731--260 
         & $5.1\times 10^{-9}$
	 & $4.3\times 10^{33}$
	 & 110 
	 & 15  	 
	 & Wijnands et al.\ (\cite{wgvm02}) :
	 & Table 1, fit 2 \\
SAX J1808.4--3658
         & $5.0\times 10^{-12}$ 
	 & $\la 1.0\times 10^{31}$
	 & --- 
	 & --- 
 	 & Campana et al.\ (\cite{campanaetal02}) :
	 & Sect.\ 2.2 \\
\cline{1-2}
\cline{3-7}
\multicolumn{7}{l}{$\!\!\!^{\ast )}$ nonredshifted }\\
\end{tabular}
\end{center}
\label{tab:SXRTs}
\end{table*}

The parameters of the selected sources
are collected in Table \ref{tab:SXRTs}.
The mean mass accretion rate $\dot{M}$ 
is evaluated as $\dot{M}=\Delta M/\Delta t$,
where $\Delta M$ is a total mass accreted over a representative
period of time $\Delta t$. Both, $\Delta M$ and $\Delta t$,
should include active and quiescent periods, although
$\Delta M$ is mainly accumulated in the outburst states.
In principle, we need $\dot{M}$ averaged
over thermal relaxation time scales, $\sim 10^4$ yrs,
while the observations provide us with sparse data over periods
not longer than several decades. 
For Aql X-1, the mean $\dot{M}$ has
been estimated by Rutledge et al.\ (\cite{rbbpz00})
(their Sect.\ 5, an estimate from the RXTE/ASM light curve history).
For Cen X-4 and 4U 1608--522, we obtain $\dot{M}$ 
from Table 9 in Chen et al.\ (\cite{csl97}).
Following these authors, we take $\Delta t=8.67$ yrs for
4U 1608--522, which is a frequently bursting source
($\Delta M$ is estimated for 6 outbursts 
in the period from 1970 to 1979). We take
$\Delta t=33.16$ yrs for Cen X-4 (with $\Delta M$ given
for the only two outbursts in 1969 and 1979), 
adding the period from 1979
till now when no outbursts were observed.
Since the active states may be very rare for this source,
it is safer to consider the obtained $\dot{M}$ as an upper limit.
KS 1731--260 recently (about 1.5 years ago) returned
into quiescence after having actively accreted
for $\approx 11.5$ years.
For this source, we take 
$\Delta t$=11.5+1.5=13 yrs. We estimate
the mass $\Delta M$ accreted during the long outburst state
from the value of the mean outburst flux given by Rutledge et al.\
(\cite{rbbpzu02}, Sect.\ 3.1). The estimation 
is made in the same manner as in 
Chen et al.\ (\cite{csl97}, Sect.\ 5.1.4). Since 
the recurrence time is unknown,
our value of $\dot{M}$ is most probably an upper limit.
Finally, $\dot{M}$ for SAX J1808.4--3658
was estimated by 
Bildstein \& Chakrabarty (\cite{bc01}) and
Campana et al.\ (\cite{campanaetal02}). 

The values of $L_\gamma^\infty$ in Fig.\ \ref{fig3}
are meant to refer to the quiescent thermal luminosity
from the NS surfaces.
For the first four sources, these values are obtained
from the values of $T_{\rm s}^\infty$ and
$R^\infty$ given in Table \ref{tab:SXRTs}. The
values of $T_{\rm s}^\infty$ and $R^\infty$
were evaluated by the authors cited in Table \ref{tab:SXRTs} by
fitting the observed spectra with the hydrogen
atmosphere models. Note that the values of the surface temperature
for Cen X-4 and 4U 1608--522 are nonredshifted.
We have redshifted them assuming
$M=1.4\, {\rm M}_\odot$ and $R=12$ km. The spectrum of SAX J1808.4--3658
in quiescence is well fitted by a power law, i.e., no
surface thermal emission has been detected.
The estimates of the upper limit of $L_\gamma^\infty$
are model dependent and range from
about $10^{30}$ erg s$^{-1}$ to
$2.5 \times 10^{31}$ erg s$^{-1}$ 
(Campana et al.\ \cite{campanaetal02}).
We take $L_\gamma^\infty = 10^{31}$ erg s$^{-1}$
with the notice that the actual surface luminosity may be 
much lower.

If the interpretation of quiescent emission as the thermal
emission from the NS surfaces is correct, then all five NSs
are heated to the neutrino emission stage 
($L_{\rm dh}^\infty > L_\gamma^\infty$).
Since $L_{\rm dh}^\infty$ is reliably determined by the 
theory (Haensel \& Zdunik \cite{hz90,hz03})
for a known $\dot{M}$, and $L_\gamma^\infty$ is measured,
one can immediately estimate the neutrino luminosity of any source
from the thermal balance, 
Eq.\ (\ref{Main}):
$L_\nu^\infty=L_{\rm dh}^\infty-L_\gamma^\infty$. 
In all our cases $L_\nu^\infty$ is
comparable with $L_{\rm dh}^\infty$ (Fig.\ \ref{fig3}).

As seen from Fig.\ \ref{fig3},
we can treat NSs in 4U 1608--52 and Aql X-1 as low-mass
NSs with very weak neutrino emission from their cores
(suppressed by strong nucleon superfluidity).
The NSs in Cen X-4 and SAX J1808.4--3658 seem to require the enhanced 
neutrino emission and are thus more massive.
The status of the NS in KS 1731--26 is less certain because
of poorly determined $\dot{M}$ (see above).
If the real value of $\dot{M}$ is close to that
in Table \ref{tab:SXRTs} it may also require
some enhanced neutrino emission.
Similar conclusions have been made by several authors
(particularly, by Ushomirsky \& Rutledge \cite{ur01};
Colpi et al.\ \cite{cgpp01};
Rutledge et al.\ \cite{rbbpz01,rbbpzu02,rbbpz02}; 
Brown et al.\ \cite{bbc02};
Wijnands et al.\ \cite{wgvm02}) 
with respect to some of these sources or selected groups.
Colpi et al.\ (\cite{cgpp01}) presented also the
heating curves for  
specific models of low-mass and high-mass 
NSs with superfluid nucleon cores and
suggested that by tuning nucleon superfluidity and 
NS masses one can explain the data.
Using the toy-model, we can present a general analysis of the problem
for different EOSs of NS interiors
(assuming, of course, that all the sources have to be interpreted
in terms of one EOS). In this way we can
quantify the assumptions on enhanced neutrino emission 
in terms of pion condensed, kaon condensed, and Durca-allowed
nucleon models of matter.

Disregarding the SAX source for the moment,
we can treat the NS in Cen X-4 either
as a high-mass NS (with a kaon-condensed or quark core) or as
a medium-mass NS (with a pion-condensed, quark,
or Durca-allowed nucleon core); thus we cannot
determine the nature of superdense matter.  
If the data on SAX J1808.4--3658 are really relevant
for our analysis we have the only choice
to treat the NS as a high-mass NS with the
nucleon core (and the NS in Cen X-4 as the medium-mass NS
with the nucleon core). This would mean that NS cores do not
contain exotic phases of matter.

Our toy model is too flexible and does not allow
us to fix the position of the transition layer in the
stellar cores where the slow neutrino emission 
transforms into the fast one.
Adopting a specific EOS of NS interiors
(with this position determined by microphysics input)
we would be able to construct the sequences
of heating curves for the stars with different $M$, and
attribute certain values of $M$ to any source
(``weigh'' NSs in SXRTs, as proposed by Colpi et al.\ \cite{cgpp01},
just as in the case of cooling isolated NSs considered, e.g.,
by Kaminker et al.\ \cite{kyg02}).
We intend to do this in our future publications,
using an exact cooling code and
taking into account some effects neglected 
in our simplified approach. 

In particular,
we will account for
the presence of light elements on the NS surfaces:
they change the thermal conductivity
of the NS heat-blanketing envelope and the relation
between the surface and internal temperature of NSs.
The effect is well known for cooling NSs
(Potekhin et al.\ \cite{pcy97}) and has been applied recently
to accreting NSs (Brown et al.\ \cite{bbc02}).

We will also carefully treat the effects
of baryon superfluidity in NS interiors
(just as for cooling NSs, see, e.g., Kaminker et al.\ \cite{kyg02}).
Particularly, we will study the effects of Cooper-pairing
neutrino emission of baryons neglected in the present analysis.
Under certain conditions, this neutrino emission would violate
our general assumptions on the neutrino emissivity
$Q_\nu(T,\rho)$ and complicate our study.
For instance, according to our estimates,
the $^3$P$_2$ neutron superfluidity 
with the maximum values of the density dependent critical temperature
$T_{\rm cn}(\rho)$ from
$\sim 10^8$ K to $\sim 2 \times 10^{9}$ K
in the nucleon NS core
would produce a powerful Cooper pairing neutrino emission
and strongly affect the thermal states of accreting NSs.
However, the same effect would initiate 
a really fast cooling
of not too massive isolated NSs in
contradiction with the observations
of isolated cooling NSs (e.g., Kaminker
et al.\ \cite{kyg02}). Thus the presence of the indicated
neutron superfluidity can be rejected on these
grounds. 

Because of the similarity between the
heating and cooling curves, the observations of cooling isolated
NSs and accreting NSs in SXRTs can be analyzed
together employing the same EOSs of NS interiors.
This increases the statistics of the sources and
the confidence of the results.
The theory of cooling NSs has recently been confronted
with observations by Yakovlev \& Haensel (\cite{yh02}).
Some cooling NSs (first of all, RX J0822--43 and
PSR 1055--52) can be interpreted 
(Kaminker et al.\ \cite{kyg02}) as low-mass
NSs with strong proton superfluidity in their cores. 
Other sources (first of all, Vela and Geminga)
seem to require a fast neutrino emission but the
nature of this emission (a choice of fast-cooling
model from Table 2) is uncertain, just as for SXRTs
disregarding the data on SAX J1808.4--3658.
In this context, the latter source is now
the only one which indicates the absence of exotic phases of
matter in NS cores.

The assumption that the observed X-ray emission
of SXRTs in quiescence emerges from the NS interior is
still an attractive hypothesis. In any case the theory
of deep crustal heating is solid and leaves no doubts
that this heating
does occur in accreting NSs leading to observational consequences.

\begin{acknowledgements}
The authors are grateful to anonymous referee
for critical comments and to J.~L. Zdunik
for new calculations of the energy release
in the deep crustal heating. 
KPL and DGY acknowledge
hospitality of N.\ Copernicus Astronomical
Center in Warsaw.
This work was supported in part by the
RBRF (grants Nos. 02-02-17668 and 03-07-90200)
and KBN (grant 5 P03D 020 20).
\end{acknowledgements}

\end{document}